# Morphological Control of Bundled Actin Networks Subject to Fixed-Mass Depletion

James Clarke[a], Lauren Melcher[b†], Anne D. Crowell[c†], Francis Cavanna[a†], Justin R. Houser[d], Kristin Graham[d], Allison M. Green[c], Jeanne C. Stachowiak[d], Thomas M. Truskett[c], Delia J. Milliron[c], Adrianne M. Rosales[c], Moumita Das[e], and José Alvarado[a]

a. UT Austin Department of Physics, 2515 Speedway, Austin, TX, USA.; b. School of Mathematical Sciences, Rochester Institute of Technology, Rochester, NY, USA; c. UT Austin McKetta Department of Chemical Engineering, E 24th St, Austin, TX, USA.; d. UT Austin Department of Biomolecular Engineering, Austin, TX, USA; e. School of Physics and Astronomy, Rochester Institute of Technology, Rochester, NY, USA

†: Authors Contributed Equally

Depletion interactions are thought to significantly contribute to the organization of intracellular structures in the crowded cytosol. The strength of depletion interactions depends on physical parameters like the depletant number density and the depletant size ratio. Cells are known to dynamically regulate these two parameters by varying the copy number of proteins of a wide distribution of sizes. However, mammalian cells are also known to keep the total protein mass density remarkably constant, to within 0.5% throughout the cell cycle. We thus ask how the strength of depletion interactions varies when the total depletant mass is held fixed, a.k.a. fixed-mass depletion. We answer this question via scaling arguments, as well as by studying depletion effects on networks of reconstituted semiflexible actin *in silico* and *in vitro*. We examine the maximum strength of the depletion interaction potential $U^*$ as a function of $q$, the size ratio between the depletant and the matter being depleted. We uncover a scaling relation $U^* \sim q^\zeta$ for two cases: fixed volume fraction $\varphi$ and fixed mass density $\rho$. For fixed volume fraction, we report $\zeta < 0$. For the fixed mass density case, we report $\zeta > 0$, which suggests the depletion interaction strength increases as the depletant size ratio is increased. To test this prediction, we prepared our filament networks at fixed mass concentrations with varying sizes of the depletant molecule poly(ethylene glycol) (PEG). We characterize the depletion interaction strength in our simulations via the mesh size. In experiments, we observe two distinct actin network morphologies, which we call weakly bundled and strongly bundled. We identify a mass concentration where different PEG depletant sizes leads to weakly bundled or strongly bundled morphologies. For these conditions, we find that the mesh size and intra-bundle spacing between filaments across the different morphologies do not show significant differences, while the dynamic light scattering (DLS) relaxation time and storage modulus between the two states do show significant differences. Our results demonstrate the ability to tune actin network morphology and mechanics by controlling depletant size and give insight into depletion interaction mechanisms under the fixed-depletant-mass constraint relevant to living cells.

## Introduction

The intracellular cytosol is a crowded environment with a high concentration of soluble proteins. Many kinds of intermolecular interactions, such as van-der-Waals, electrostatic, and hydrophobic interactions ultimately determine protein structure, binding, and function. In addition, steric effects play a significant role and contribute to intermolecular interactions in complex ways: they mediate short-range repulsion, promote nematic alignment and defects, and mediate interactions with boundaries and confining geometries [1]. Strikingly, a crowded solution of small depletant molecules can, through steric repulsion, drive short-ranged attractive interactions (depletion interactions) between larger, μm-scale structures [2]. Depletion interactions from intracellular proteins are thus believed to contribute to the spatial organization of intracellular structures such as the nucleus, amyloid fibrils, and the cytoskeleton [3]. However, quantifying how depletion interactions contribute to intracellular organization remains a challenging task, given the large number of short-ranged steric interactions and complex spatial hierarchy in living cells.

Although depletion in living cells remains difficult to characterize, researchers have successfully studied depletion in a wide range of model systems: colloids [4], polymer-coated nanocrystals [5], actively stirred rods and beads [6], actin and DNA polymers in solutions [7], DNA polymers suspended with increasing sizes of dextran molecules [8], and microtubules driven by poly(ethylene glycol) (PEG) polymers [9]. These studies have found that the maximum strength of the depletion interaction $U^*$ at close contact scales with the overlap volume $V_{overlap}$, that is, the additional volume made available to the depletants when the colloids are in close contact. It has been well established that increasing depletant concentration generally increases the depletion interaction strength and increasing depletant size generally increases the depletion interaction length scale. This has been determined typically by performing many experiments which systematically vary depletant concentration and depletant size.

However, cells are not thought to significantly vary protein concentration. In fact, mammalian cells have been recently shown to keep the total protein mass density remarkably constant, to within 0.5% throughout the cell cycle [10]. Instead, cells dynamically regulate the copy number of proteins over a wide distribution of sizes. This raises the following question: how are depletion interactions affected when the total depletant mass is held fixed but the depletant size is allowed to vary? That is, what is $U^*(q)$ for fixed mass density $\rho$? Here, $q$ is the size ratio between the depletant and the matter being depleted. Surprisingly, however, prior studies on depletion interactions do not appear to provide a clear answer to this question. One reason for this is that depletant concentration can be given in terms of number density, volume fraction, or mass density, and these three quantities appear to be roughly equivalent measures. Another reason is that, experimentally, varying depletant concentration or depletant size *in situ* remains technically challenging. Answering this question is essential to better understand how steric repulsion and depletion interactions contribute to intracellular structures in crowded environments. In addition, understanding and characterizing fixed-mass depletion could be leveraged to develop novel biomimetic materials. Control over depletant size distribution could be a useful experimental knob for affecting interactions, structure, and mechanics of novel materials for biomedical or soft-robotics applications without requiring direct covalent, or specific interactions between constituent polymers or colloids.

In order to study fixed-mass depletion, we turn to a combination of scaling arguments, simulations, and experiments that utilize an *in vitro* model system of actin biopolymers and PEG depletants. Actin is one of the most prevalent biopolymers in the cell and a key component driving mechanics in the cellular cortex. It exhibits a wide polymorphism in its ability to polymerize and then form higher-order structures out of the polymerized actin [11]. Of specific interest for the work presented herein are actin bundles, which are utilized in lamellipodia [12], filopodia [11], stress fibers [13], and microvilli [14]. The mechanisms driving bundle formation are varied, and several have been identified *in vitro*. Electrostatic bundling via counterion condensation screens charges held on the actin, and by electric attraction pulls actin filaments into bundles [15–17]. Physiological bundling depends on specific proteins that are known to specifically bind to and crosslink actin [18], such as fascin [19], α-actinin [20], and filamin [21,22]. Bundle formation can also be caused by the depletion interaction. Above a threshold concentration of depletant molecules, individual actin filaments self-bundle into rings and raquets [23], and more concentrated networks of actin filaments form networks of entangled bundles [24,25]. We elect to use PEG as a depletant. PEG is known to bundle actin via a depletion interaction and has an easily tunable molecular weight [24,26]. In the second half of the work, we characterize two distinct states of our system, $\Gamma_{6k}$ and $\Gamma_{20k}$, that demonstrate a large change in morphology when depletant mass distribution is changed slightly from 6 kDa to 20 kDa under the fixed-mass constraint. In particular, structural measures such as the mesh size, the spacing between bundles, the relaxation time of the network, and the bulk mechanical properties of the network are investigated.

## Results

**Analytical scaling arguments: fixed-volume vs. fixed-mass constraints.** We begin by developing an analytical framework to investigate fixed-mass depletion. In order to understand how the strength of the depletion interaction varies while depletant mass is held fixed, we examine here the seminal Asakura-Oosawa model [27] for two cases: fixed volume fraction $\varphi$ and fixed mass density $\rho$. We first consider two colloidal spheres of fixed diameter $\sigma$, immersed in a solution containing spherical depletants (e.g. polymers) with variable diameter $\tau$ and variable number density $n$. The interaction potential $U$ between the two colloidal spheres is given by:

$$U(d) = -nkT V_{overlap}(d) \qquad d \leq 1 \tag{1}$$

where $k$ is the Boltzmann constant, $T$ the temperature, $V_{overlap} = \frac{\pi}{6}\sigma^3(1+q)^3\left(1 - \frac{3}{2}d + \frac{1}{2}d^3\right)$ the overlap volume [28], $D$ the distance between the two colloids' center of masses, $d = \frac{D}{\sigma(1+q)}$ the distance normalized by the maximal extent of the depletion interaction, and $q = \frac{\tau}{\sigma}$ the size ratio. We wish to determine the maximal value $U^*$ of the interaction potential, which occurs when the two colloids are in contact and thus $D = \sigma$, or equivalently $d = \frac{1}{1+q}$. Furthermore, we wish to determine how the scaling relation $U^* \sim q^\zeta$ is affected by the two cases of fixed $\varphi$ and fixed $\rho$.

We now consider the above expressions for the scenario where the depletant is much smaller than the colloids. We only consider values of $q$ smaller than 0.1547 because the Asakura-Oosawa model breaks down at higher values of $q$ [28]. In this regime, we note that the right-hand factor containing the third-order polynomial in $d$ in the above expression for $V_{overlap}$ Taylor expands to scale as $q^2$, yielding $V_{overlap} \sim \sigma^3 (1+q)^3 q^2$. For the fixed volume fraction case, we have $\varphi = \frac{4}{3}\pi \left(\frac{\tau}{2}\right)^3 n$. Substituting into the above equation for $U(d)$ yields

$U^* \sim -\varphi \left(\frac{1+q}{q}\right)^3 q^2 \sim q^{-1}$. The negative scaling exponent $q^\zeta \equiv q^{-1}$ tells us that breaking up a fixed volume of spherical depletants into smaller spheres increases the number density to allow the interaction potential to diverge as $q \to 0$. Thus, for fixed volume fraction, smaller depletants result in stronger depletion interactions.

In contrast, for the fixed mass density case, we have $\rho = MW\, n \sim R_g^{\frac{5}{3}} n$, where we use the Flory scaling relation for polymers [29]. Substituting into the above equation for $U(d)$ produces a positive scaling exponent $\zeta$:

$$U^* \sim -\rho \sigma^{\frac{4}{3}} \frac{(1+q)^3}{q^{5/3}} q^2 \sim q^{\frac{1}{3}}. \tag{2}$$

This relation predicts that two colloidal spheres will experience a stronger attractive interaction if polymer depletants become larger while maintaining mass density fixed [28].

To extend this analysis to actin, we turn to the case where we consider two cylindrical rods to better approximate actin filaments. The maximum depletion interaction potential between a pair of non-parallel cylindrical rods of cross-sectional diameter $\sigma$, is given by $\frac{U^*}{kT} = -\frac{\pi}{2} n \sigma \tau^2 = -\frac{\pi}{2} n q^{-1} \tau^3$ [30]. Thus, at fixed volume fraction, we have $\varphi = \frac{4}{3}\pi\left(\frac{\tau}{2}\right)^3 n$, which gives an interaction potential

$\frac{U^*}{kT} = -3\varphi q^{-1}$.

Here we observe a negative scaling exponent $\zeta = -1$ as before. For the fixed mass density case we have $\rho = M_w n \sim \tau^{\frac{5}{3}} n$, whose associated interaction potential is given by

$$\frac{U^*}{kT} = -\frac{\pi}{2}\sigma^{\frac{4}{3}} \rho q^{\frac{1}{3}}, \tag{3}$$

with $q = \frac{\tau}{\sigma}$ as defined earlier. We observe the same positive scaling exponent $\zeta = \frac{1}{3}$ in the interaction potential as obtained for the pair of spheres described above.

Further, for the case of parallel cylinders, with interaction potential $\frac{U^*}{kT} = -\frac{2\sqrt{2}}{3} L n \tau^{\frac{3}{2}} \sigma^{\frac{1}{2}}$ [30] we obtain

$$\frac{U^*}{kT} = -\frac{4}{\pi} L \sigma^{-1} q^{-\frac{3}{2}}, \tag{4}$$

for the case of fixed volume fraction, and

$$\frac{U^*}{kT} = -\frac{2\sqrt{2}}{\pi} L \sigma^{\frac{2}{3}} q^{\frac{1}{6}}, \tag{5}$$

for the case of fixed mass density. With parallel cylinders, we get different scaling relationships of $\frac{U^*}{kT}$ with $q$ but obtain a similar reversal of sign of the scaling exponent $\zeta$ as was observed for the case of spherical particles and non-parallel cylinders. Namely, the interaction potential decreases with $q$ for fixed volume fraction and increases with $q$ for fixed mass density.

Note that these relationships only hold when $\frac{\pi n \tau^3}{6} \ll 1$. For very dense systems where this condition is violated, we can apply Carnahan-Starling theory such that the number density of depletants $n$ should be replaced by $nZ$, where $Z = \frac{1+\eta+\eta^2-\eta^3}{(1-\eta)^2}$, and $\eta = \frac{\pi n \tau^3}{6}$ [30]. For a fixed volume fraction $\varphi$, $Z = \frac{1+\varphi+\varphi^2-\varphi^3}{(1-\varphi)^2}$, i.e. there is no change to the scaling of $\frac{U^*}{kT}$ with q. However, for a fixed mass density, $Z = 1 + \frac{1}{2}\pi\rho\sigma^{4/3}q^{4/3} +$ higher order terms, which will change the scaling when $q \sim 1$.

Given these results, we hypothesize that the depletion interaction strength increases as the depletant size is increased, subject to the constraint that mass is held fixed. In order to test this hypothesis, we turn to numerical simulations.

**Validation via simulation: larger depletants increase actin bundling.** To investigate the underlying depletion mechanism for fixed depletant mass systems we developed a numerical model, which consists of actin filaments interacting with spherical PEG particles in three dimensions. Actin filaments are long compared to the PEG molecules. The persistence length of actin is 17 μm [31,32] and the diameter of F-actin is about 7 to 8 nm [33,34]. The radius of gyration for PEG is given by $R_g \sim MW^{3/5}$ [29]. In this section we examine PEGs with $R_g$ both smaller and larger than the diameter of F-actin. Given the difference in length scales and sufficient concentrations of PEG, there is an entropic favorability for PEG to occupy maximal volume. PEG accomplishes this by depleting the distances between actin filaments in its local environment, thus inducing bundle formation. When two filaments are closer than twice the radius of a PEG molecule, the filaments act as a semi-permeable structure where fluid is permitted to flow, but PEG particles are excluded from the region. This attracts the two filaments toward one another, effectively depleting any separation between them [35].

To model this type of interaction numerically, each actin filament is modeled as a chain made of beads, and the relative cost of stretching and bending the chain is informed by known mechanical properties of actin filaments. We model the pairwise interactions between any two particles, whether belonging to actin bead-chains, or PEG, or between actin and PEG using a Lennard Jones potential, $V_{LJ} = 4\epsilon\left[12\frac{\sigma_{ab}^{12}}{r_{ij}^{13}} - 6\frac{\sigma_{ab}^{6}}{r_{ij}^{7}}\right]$, where $\sigma_{ab} = \frac{\sigma_a + \sigma_b}{2}$, where $\sigma_a$ and $\sigma_b$ represent the diameters of two reference beads. The interparticle distance between two reference particles with indices $i$ and $j$ is denoted by $r_{ij}$ and the strength of the interaction's potential is defined by $\epsilon$. The interaction potential is truncated to incorporate only repulsive interactions representing a hard core for actin-PEG interactions, while for PEG-PEG and actin-actin interactions we additionally allow attractive interactions to mimic the Van-der Waals interactions.

The motion of the PEG particles follows the overdamped Langevin equation, $\frac{dr}{dt} = \frac{D}{K_b T}F_{LJ}\sqrt{2D\eta}$, where the interparticle interaction force $F_{LJ}$ is derived from the Lennard Jones potential defined above, and the thermal diffusion is represented by the noise term $\sqrt{2D\eta}$, where $D$ is the diffusion constant and $\eta$ is a vector drawn from a Gaussian distribution with a mean of 0 and variance of 1; $K_b$ and T are the Boltzmann constant and room temperature. The motion of an actin bead in strand $i$ is also given by an overdamped Langevin equation: $\frac{dr_i}{dt} = \frac{D}{K_b T}\left(\sum F_{LJij} + \sum F_{LJik} + \sum F_{LJil}\right) + \frac{D}{K_b T}\left(\sum F_{s,ij} + \sum F_{b,ij}\right) + \sqrt{2D\eta}$, where $F_{LJij}$, $F_{LJik}$, and $F_{LJil}$ are the Lennard Jones interparticle forces, and indices $i$ and $j$ describe intra-strand pairwise interactions, $i$ and $k$ describe inter-strand interactions, and $i$ and $l$ describe interactions between actin and PEG particles. The actin filaments resist stretching (or compression) and bending with forces $F_{s,ij}$ and $F_{b,ij}$, respectively, obtained from the corresponding stretching and bending deformation energies, $U_s = \frac{K_s}{2}(l-l_0)^2$ and $U_b = \frac{K_b}{2}(\theta-\theta_0)^2$. In the equations above, $l$ represents the distance between two nearest-neighbor actin beads in a strand, and $l_0$ describes the equilibrium rest length for actin beads which is given by the actin bead diameter. The stretching stiffness is $K_s$, the bending rigidity is $K_b$ and $\theta$ is the angle between three sequential actin beads and the equilibrium value of this angle, $\theta_0$, is set to $\pi$.

We solve both Langevin equations using the Forward Euler-Maruyama approach. The timestep $dt$ is set to $10^{-3}$ and each simulation was run up to $500\,\tau$, so the system had enough time to equilibrate and reach a steady state. Simulation results are shown below in Figure 1. With our model, we present two sets of simulation cases. In the first case, we vary both the radii of the PEGs and the number density of PEGs (Figure 1a). In a second set of simulations, the PEG mass density is kept constant and the radii of gyration ranges from 4 nm to 20 nm (Figure 1b).

To explore the tradeoff associated with holding the total depletant mass fixed, while varying the size of the depletant we select 3 radii of interest $R_{g,1}$, $R_{g,2}$, and $R_{g,3}$ that correspond to ~4 nm, 8 nm, 20 nm PEGs, respectively. The simulation depletant sizes were chosen to explore a large $q$ regime. For $R_{g,1}$, the number of PEG molecules was set to 6000, the case for $R_{g,2}$ had 1890 PEG molecules, and $R_{g,3}$ had a total of 410 molecules, to keep the mass density fixed. These correspond to PEGs of MW ~20 kDa, ~70 kDa, and ~317 kDa, respectively. Statistical results corresponding to depletion interactions simulated for these different cases are shown below in Figure 1b.

The results of Figure 1b are qualitatively consistent with our predictions from the scaling arguments presented in the previous section, as we see a steady increase in the network mesh size as a result of increasing the PEG molecular weight at fixed mass density. We take the mesh size, $\xi$, to directly reflect differences in $U^*$ because increases in mesh size correspond to more effective bundling in the system [36], and thus a stronger interaction potential (see Figures S1, S2). Details on the mesh size algorithm can be found in the supplemental information (see Figure S1). Given the simulation results, we decided to test if the numerically supported scaling predictions extend to the same system studied *in vitro*.

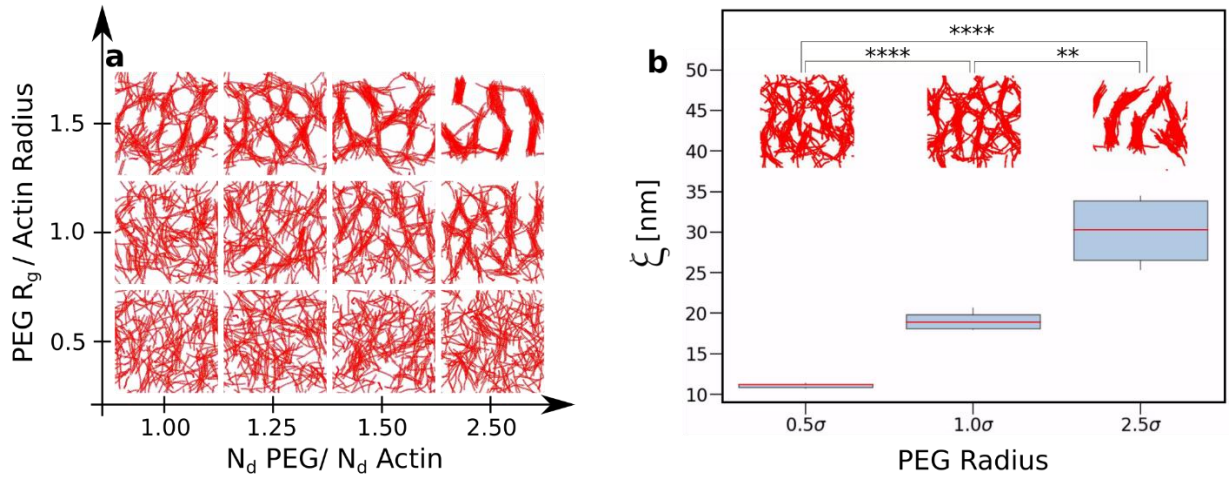

Figure 1: (a) Simulation snapshots showing the simulated equilibrium morphologies of the networks and bundles of actin filaments in the presence of PEG. These snapshots demonstrate the role of PEG molecular weight and concentration on the actin filament bundling. As the molecular weight, and thus size ratio *q*, increases (y-axis), significant bundling of actin filaments is observed. Similarly, increasing the number density, $N_d$, of PEG (x-axis) also leads to appreciably more pronounced bundles. $R_g$ is the radius of gyration for a given PEG molecule where $R_g \sim MW^{3/5}$. (b) Box plots presenting the mesh size of simulated actin gels when PEG size is varied while keeping the total mass of PEG in the system fixed, i.e. reducing $N_d$ of PEG when increasing molecular weight (N=5 simulations). Inset tiles are real space representations of network morphology for the corresponding distribution directly below them. For interpreting the boxplots - the top and bottom of each box gives the upper and lower quartile where the box height represents 50% of the distribution, the lines above and below the box represent the upper and lower whisker where the distance between represents 100% of the distribution, the red line inside the box indicates the median value. $\sigma = 8\,\text{nm}$.

**Validation via experiment: larger depletants increase actin bundling.** Given the predictions made from the modelling, we moved to study these systems with *in vitro* experimental assays. Bundled actin networks were assembled (see Methods), and subsequently imaged with a confocal microscope. We study PEGs of variable molecular weight and size: for 20kDa, $R_g \simeq 3.9\,\text{nm}$, for PEG 6kDa, $R_g \simeq 1.9\,\text{nm}$, and for PEG 2kDa, $R_g \simeq 0.9\,\text{nm}$ [29]. The primary results of this section are captured in Figure 2. Figure

2a shows confocal micrographs of fluorescently tagged actin filaments that capture our experimental parameter space, representing 36 distinct samples to give 3 replicates per sample condition. For a representation of Figure 2a in number density space see Figure S3 in the supplemental information. Figure 2b shows the results of our mesh size analysis performed on the data represented in Figure 2a.

We find qualitative agreement between the configuration space diagram produced by the simulation (cf. Fig. 1a) and that of experiment Figure 2a. In both simulation and experiment, the actin network exhibits increased bundling as either PEG concentration or PEG molecular weight are increased. However, the way the system changes to a purely bundled state appears to be more abrupt in the experiment than in simulation (see Discussion). Further, we note that we do see statistically significant differences in mesh size amongst the 1.0% wt/vol assays. This result lends experimental support to our earlier hypothesis from the simple scaling model, as well as our observed trend in simulations (cf. Fig. 1b).

We also observe two morphologically distinct regions of actin networks by systematically varying the molecular weight and concentration of PEG molecules in the assay. In Figure 2a, one can observe a distinction between weakly bundled and strongly bundled networks to the left and right of the blue line in Figure 2, respectively. In the weakly bundled regime, coexistence of actin bundles and single-filament actin networks are observed, whereas for strongly bundled samples a network of actin bundles is formed. To quantify the regions between weakly bundled and strongly bundled networks, we developed a separate image processing algorithm, the degree of bundling, that quantifies the presence of bundles and hence the observed morphologies (see supplemental information). We find that strongly bundled networks occur for a degree of bundling greater than $\sim 0.2$ [a.u.] (see Figure S4). Figure 2b demonstrates that we observe increasing bundling strength as we increase the PEG concentration, as well as for increasing PEG molecular weight. In all, we take the experimental results of Figure 2 to be in qualitative agreement with our earlier analytical and numerical predictions.

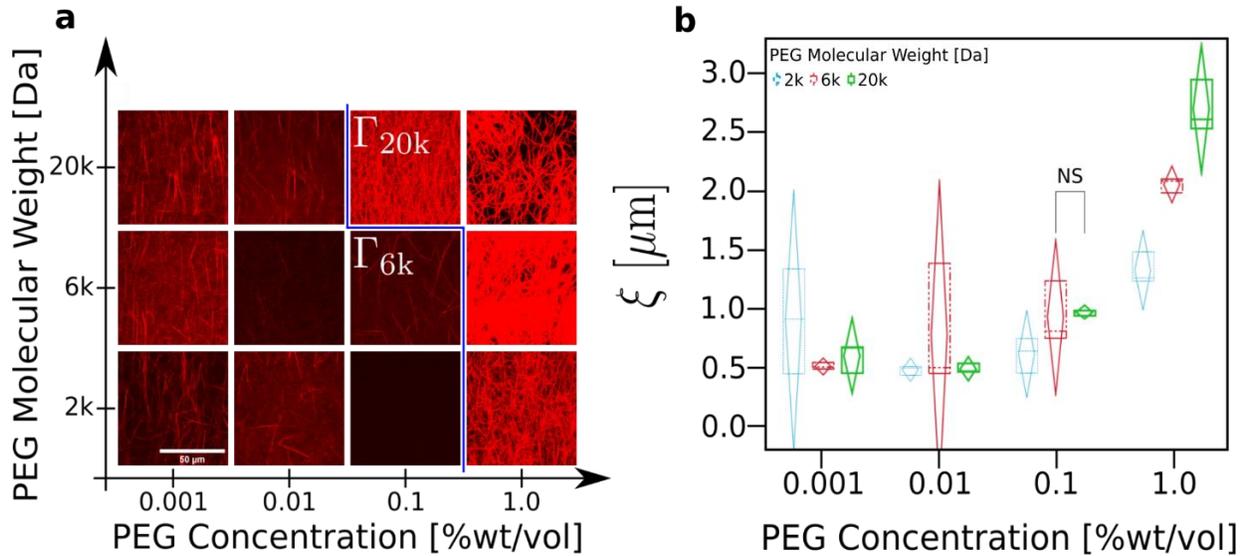

Figure 2: (a) A representative phase diagram of [actin] = 12 μM with various concentrations [%wt/vol] of PEG on the x-axis and the molecular weight of the PEG on the y-axis corresponding to a total of 36 distinct samples, giving 3 replicates for each condition. Scale bar = 40 μm . The threshold concentration of bundling (blue line) for PEG 6k and PEG 2k is at [PEG] = 1.0% (w/v), and for PEG 20k at [PEG] = 0.1% (w/v). $\Gamma_{6k}$ & $\Gamma_{20k}$ denote regions of interest for further characterization. (b) Quartile box plots with confidence diamonds of each network's mean mesh size in microns extracted from confocal micrographs for various concentrations of PEG (N=3 experiments per each sample condition). The molecular weight of the PEG is given in the inset legend where blue fine-dashes, red irregular-dashes, and green solid-line correspond to PEG at molecular weights of 2k, 6k, and 20k, respectively. Direct comparison is drawn between $\Gamma_{6k}$ & $\Gamma_{20k}$ where we can see that the mesh size is statistically identical for the two populations. NS denotes that the two populations are statistically indistinguishable. For interpreting the boxplots - the top and bottom of each box gives the upper and lower quartile where the box height represents 50% of the distribution. The horizontal line inside the box indicates the median value of the distribution. The top and bottom of the diamond are a 95% confidence interval for the mean. The middle of the diamond is the sample average, which is an estimate of the population mean.

In the remaining parts of the Results section, we characterize microscopic and bulk properties of two conditions. When we vary *q*, we observe the most dramatic change in morphology between PEG MWs 6k and 20k at [PEG] = 0.1% (w/v) which we refer to as $\Gamma_{6k}$ and $\Gamma_{20k}$, respectively. These points are shown in overlay in Figure 2a. We note that other regions of our parameter space were not studied further in favor of investigating the $\Gamma_{6k}$ and $\Gamma_{20k}$ conditions more thoroughly as we hypothesize that network properties will be distinct and may be leveraged for soft material applications. In Figure 2b, we observe that the mesh size distributions for the $\Gamma_{6k}$ and $\Gamma_{20k}$ states are statistically indistinguishable from one another due to the large variance in the $\Gamma_{6k}$ mesh size. This spread and its impact on proximity to significant separation are within our expectations for this system as these samples are located near different morphological regimes, weakly and strongly bundled, and are thus more sensitive to sample-to-sample variation. This variation manifests here in the mesh size distribution for $\Gamma_{6k}$. We begin the remaining materials characterization with an investigation into the dynamics of our actin gels using dynamic light scattering.

**Material characterization: diffusion of actin polymers in Dynamic Light Scattering.** To determine the diffusive properties of $\Gamma_{20k}$, $\Gamma_{6k}$, and a control without PEG, the samples were probed with a Zetasizer Nano ZS instrument, with a *q*-vector of $|q| = \frac{4\pi \cdot 1.334}{632\,\text{nm}} \sin(173°/2) = 26.4\,\mu m^{-1}$ (See Methods). This *q*-vector was chosen to set the scale of the dynamics probed within the network.[37] Correlation curves describing the timescale associated with decorrelation of the system from an original scattering state for each condition measured with dynamic light scattering are represented (see Figure S5). The DLS correlation curves are fit using Eq. 3 (see Methods) to extract relevant parameters, relaxation time, $\tau_f$, and stretch exponent, $\beta$. The results are shown below in Figure 3.

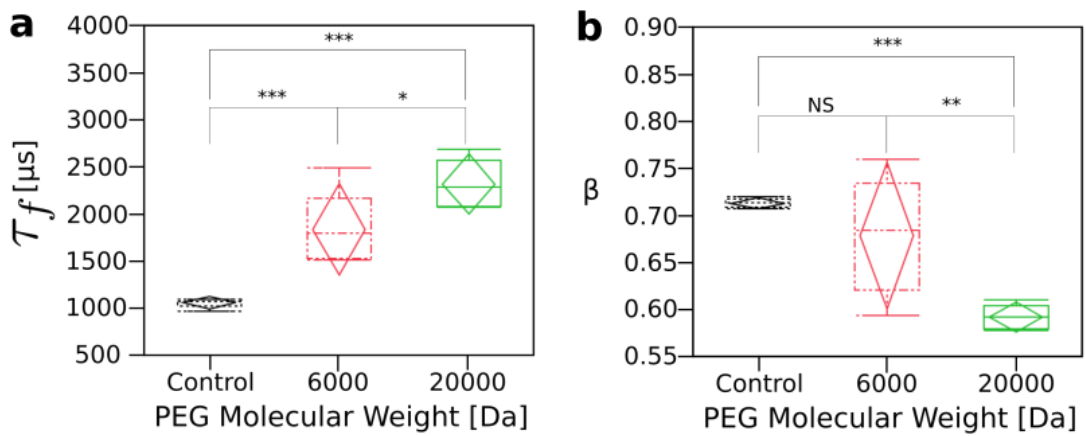

Figure 3: Each DLS correlation curve (See Figure S5) was fit, and a relaxation time, $\tau_f$, and stretch exponent, β, were obtained from the fit. The relaxation time for each condition is represented in (a) and was $\tau_{f,20} = 2314 \pm 40\,\mu s$, $\tau_{f,6} = 1833 \pm 22\,\mu s$, and $\tau_f = 1053 \pm 3\,\mu s$. The relaxation time for the $\Gamma_{20k}$ condition was significantly different from the relaxation time for the $\Gamma_{6k}$ condition (p = 0.0166), and the $\Gamma_{6k}$ condition was significantly different from the state with no PEG, (*p* = 0.0007). The stretch exponent is represented in (b) and is $\beta_{20} = 0.59 \pm 0.01$, $\beta_6 = 0.68 \pm 0.06$, and $\beta_c = 0.713 \pm 0.004$. The difference between relaxation times for the control and the $\Gamma_{6k}$ condition is not significant (*p* = 0.1668), but the difference between the $\Gamma_{6k}$ condition and the $\Gamma_{20k}$ condition is significant ($p = 0.0030$). For interpreting the boxplots - the top and bottom of each box gives the upper and lower quartile where the box height represents 50% of the distribution, the horizontal lines above and below the box represent the upper and lower whisker where the distance between represents 100% of the distribution. The horizontal line inside the box indicates the median value of the distribution. The top and bottom of the diamond are a 95% confidence interval for the mean. The middle of the diamond is the sample average, which is an estimate of the population mean.

Actin-PEG systems polymerize on the order of tens of seconds, in contrast to systems which experience gelation over dozens of minutes [38,39]. Consequently, we observe no polymerization phase, and the plots represent the system at or close to a stabilized state. The relaxation time of the $\Gamma_{20k}$ condition, $\tau_{f,20} = 2314 \pm 40 \mu s$, is roughly twice that of the control condition, $\tau_f = 1053 \pm 3 \mu s$, with the $\Gamma_{6k}$ condition somewhere in between, at $\tau_{f,6} = 1833 \pm 22 \mu s$. Relaxation time corresponds to the time needed for the intensity correlation function to decay. Longer times correspond to less diffusive networks, and shorter times correspond to networks undergoing more motion. In addition, fitting the correlation function yields the stretching exponent $\beta$, which indicates how broad the relaxation times are. $\beta$ is defined between 0 and 1, and $\beta = 1$ corresponds to a single relaxation time, while $\beta < 1$ corresponds to an increased broadening of relaxation times as $\beta$ decreases to 0. The stretched exponential fits reveal the $\Gamma_{20k}$ condition has a lower stretching exponent $\beta$ than that of $\Gamma_{6k}$ and the control, which suggests a more heterogenous distribution of relaxation timescales for the $\Gamma_{20k}$ condition than the $\Gamma_{6k}$ condition. Stretched exponent and relaxation time calculations have previously characterized formation of colloidal gels [38] and precursors to contraction of active networks [40].

**Material characterization: intra-bundle distance using FRET.**

The $\Gamma_{6k}$ and $\Gamma_{20k}$ states were also characterized using Förster Resonant Energy Transfer (FRET). FRET has been proposed as a sensing mechanism in a myriad of biochemical contexts [41] and utilized to study membrane crowding and steric pressure on membranes [42,43]. In this study we utilized FRET as a sensor to measure the distribution in intra-bundle distances between neighboring actin filaments in our system, where perturbations in the fluorescent lifetime can be correlated to real-space distances between a donor-acceptor pair [44].

Using the fluorescence lifetimes measured in both $\Gamma_{6k}$ and $\Gamma_{20k}$, $\tau_{DA}$, along with the fluorescence lifetime of the sample without any acceptor fluorophores, $\tau_D$, we calculate the FRET efficiency $E$ with,

$$E = 1 - \frac{\tau_{DA}}{\tau_D} \tag{6}$$

Given that this is measured on a per-pixel basis, with spatial separation of 300 nm per pixel in the sample, the measured efficiency is built using the mean of the respective fluorescence lifetimes, giving a spatially averaged efficiency, $\langle E \rangle$. With this, the average donor-acceptor distance is calculated using,

$$R_{DA} = R_0 \left( \frac{1}{\langle E \rangle} - 1 \right)^{\frac{1}{6}}, \tag{7}$$

where $R_0$ is the Förster radius determined to be 6.72 nm by the choice of donor and acceptor in the system [44].

The main result of the FRET analysis is that the intra-bundle spacing between filaments is statistically indistinguishable across our samples as demonstrated in Figure 4 [45].

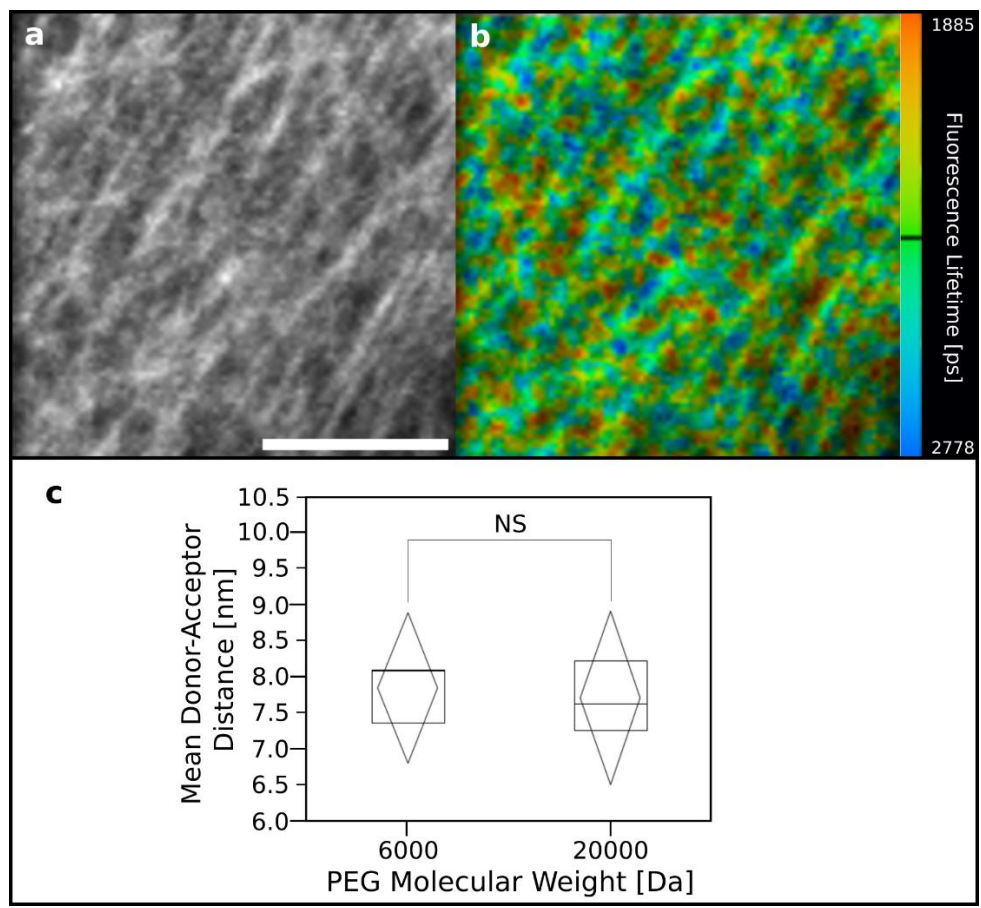

Figure 4: (a) FRET intensity confocal micrograph of a $\Gamma_{6k}$ bundled actin sample. Scale bar = 50 μm. (b) FRET fluorescence lifetime overlay of micrograph shown in a. Color bar represents the distribution in fluorescence lifetimes measured in the sample. The solid black line represents the mean of the distribution. (c) Quartile box plots and confidence diamonds for mean donor-acceptor distance giving the mean intra-bundle spacing for $\Gamma_{6k}$ and $\Gamma_{20k}$, NS denotes that the two distributions are statistically equivalent. For interpreting the boxplots - the top and bottom of each box gives the upper and lower quartile where the box height represents 50% of the distribution. The horizontal line inside the box indicates the median value of the distribution. The top and bottom of the diamond are a 95% confidence interval for the mean. The middle of the diamond is the sample average, which is an estimate of the population mean.

**Material characterization: bulk mechanical properties of bundled actin networks.** Previous studies characterized the bulk rheological properties of actin networks bundled via physical crosslinks, such as scruin [46,47] as well as without physical crosslinks [48]. The rheological properties of actin networks bundled via depletion forces have also been studied with respect to changes in PEG concentration [26]. As previously noted, however, the effect of PEG molecular weight on the bulk rheological properties of bundled actin networks has not been quantified. Thus, in this study, oscillatory shear rheometry is used to characterize the bulk properties of the bundled actin networks. Specifically, we measure the storage modulus ($G'$) and loss modulus ($G''$) as a function of time for the $\Gamma_{6k}$ and $\Gamma_{20k}$ states (Figure 5). $G'$ is especially of interest because it provides a measure of the bulk elasticity of the networks. Three replicates of each network type were created in pairs: for each replicate, enough actin for two networks was prepared, and from this actin preparation, one $\Gamma_{6k}$ sample and one $\Gamma_{20k}$ sample were each prepared. This experimental design was then executed on two different geometries, parallel plate (PPG) and conical (CG). Figure 5a shows the evolution of $G'$ and $G''$ over time for one of the CG replicates. $G'$ increases during the first 10 minutes for both networks. Because the components of the network are mixed immediately before beginning measurements on the rheometer, the increase in $G'$ during that period is attributed to the formation of the bundled actin networks, corresponding to the so-called gelation time scale [39]. After this time, $G'$ plateaus to a steady value, suggesting that network formation is complete. The networks were also characterized through frequency sweeps. Figure 5b shows the frequency sweeps performed on the same replicate as in Figure 5a. $G'$ and $G''$ both exhibit frequency-dependent behavior, and at all frequencies, $G'>G''$. In addition, strain sweeps on the formed networks can be seen in Figure S6. Figure 5c shows $G'$ from the plateau in the time sweeps for all CG replicates. A representative evolution of $G'$ for PPG is shown in the supplemental information (see Figure S7). Notably, across all replicates $G'$ is higher for $\Gamma_{20k}$ than for $\Gamma_{6k}$, suggesting that a stiffer network is formed when a higher molecular weight PEG chain is used. When comparing PPG and CG distributions in $G'$ we observe that they are statistically different via t-test with *p*-values of 0.05 and 0.03 for $\Gamma_{6k}$ and $\Gamma_{20k}$, respectively. This difference is attributed to the onset of inertial effects for the PPG, indicated by a phase angle greater than 170 degrees [49]. In Figure 5d, we normalize within each CG replicate to the equilibrium value of $\Gamma_{6k}$. This normalization procedure eliminates systematic variation associated with pipetting variance between replicates. With this analysis, we find that $\Gamma_{20k}$ replicates are stiffer than $\Gamma_{6k}$ replicates.

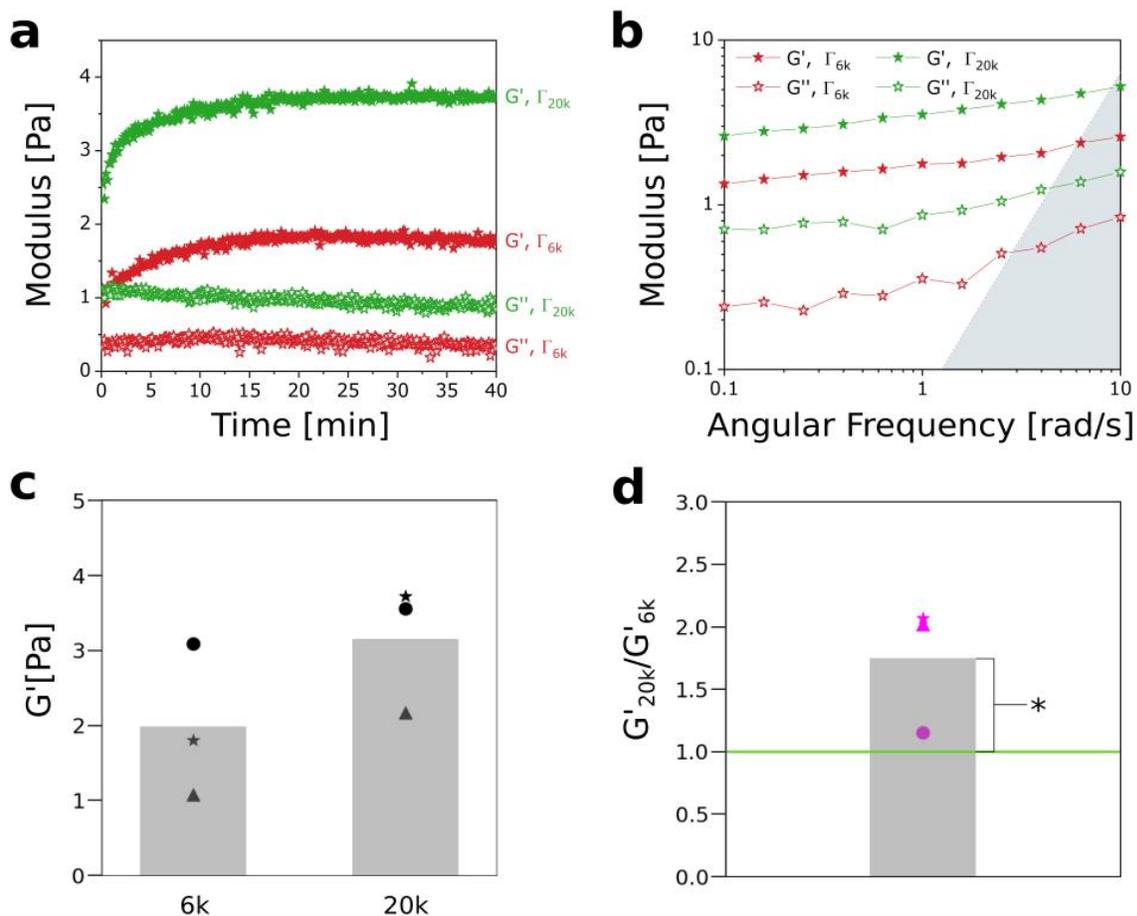

Figure 5: (a) Time sweeps showing the evolution of the storage moduli, $G'$ (closed stars) and loss moduli, $G''$ (open stars) of the bundled actin networks formulated with $\Gamma_{6k}$ (red) and $\Gamma_{20k}$ (green). (b) Frequency sweeps performed at 2% strain measured $G'$ (closed stars) and $G''$ (open stars) of the bundled actin networks formulated with $\Gamma_{6k}$ (red) and $\Gamma_{20k}$ (green). The shaded region indicates conditions under which inertia is expected to dominate the measured torque. (c) Bar graphs for conical geometry measurements, with each replicate represented by a different shape. These distributions are not statistically significantly different from one another. (d) Bar graph of normalized storage moduli for all replicates for the conical geometry data. Distinct replicates are shown using shapes, with conical geometry shapes matching panel c. Z-score comparison assuming data is normally distributed gives a significant difference between $G'_{20k} / G'_{6k}$ and unity ( $G'_{6k} / G'_{6k}$ ), indicated by the green line, with a *p*-value of 0.04 (*) for the conical geometry, and < .00001 (****) for the parallel plate geometry.

## Discussion

**Fixed-mass depletion in relation to living cells.** We sought to understand how the strength of depletion interactions varies when the total depletant mass is held fixed, a.k.a. fixed-mass depletion. By framing this inquiry in the context of the Asakura-Oosawa model we are able to show that the interaction potential scales as $q^\zeta$. For the fixed volume fraction case we report values of $\zeta = -1$ for spheres and non-parallel rods, and $\zeta = -\frac{1}{2}$ for parallel rods, showing that smaller depletants therefore increase the depletion interaction strength. Meanwhile, for the fixed mass density case, we saw the sign of $\zeta$ flip such that that $\zeta = \frac{1}{3}$ for spheres and non-parallel rods, and $\zeta = \frac{1}{6}$ for parallel rods. Thus, when PEG depletant mass density is held fixed, the depletion interaction strength increases, rather than decreases.

Our simulations are qualitatively consistent with the scaling prediction for fixed mass density, although differences between the two make a direct quantitative comparison difficult for a few reasons. First, the range of $q$ between scaling model and simulation do not overlap. Strictly speaking, the AO model only holds for $q < 0.1547$ [28]. In our simulations we explore depletant that extends beyond this threshold (cf. Fig.1). However, we still observe the same expected trend given by the scaling arguments - interaction strength increases as we increase PEG molecular weight. Second, we do not have a reliable quantitative measure of $U^*(q)$ in simulation, though the presence of thicker bundles and larger mesh sizes are indications of stronger depletion interactions (cf. Fig. 1). Thus, although we cannot experimentally access the exact magnitude of $\zeta$ in simulation, we can determine its sign to be positive. Third, our analytical model doesn't account for the semiflexible nature of the polymers. Despite this, we suspect that the key result pertaining to fixed mass-density depletion could extend to a broader class of semiflexible polymers as evidenced by the results of our simulation – which do consider actin filaments as semiflexible polymers. Fourth, simulations include Lennard Jones interactions, which are absent from the Asakura-Oosawa model as well as in our scaling argument. Quantifying how these effects contribute to the interaction potential in the context of fixed-mass depletion would be an interesting area of further study.

We corroborate our predictions from simulation via experiment - capturing a morphological difference between weakly bundled and strongly bundled states by systematically varying PEG molecular weight and concentration. We observe that the morphology of the network becomes increasingly bundled along both axes (cf. Figure 2). However, we find one difference between simulation and experiment, namely the nature of how the system varies across single filament, weakly bundled, and strongly bundled regimes. In the model, we observe a more gradual change, whereas in experiment the change from one morphology to the next is much more distinct. One plausible cause is finite size effects; while the simulation study consists of 200 actin filaments, the experiments study systems have $O(10^3)$ actin filaments. Furthermore, for computational simplicity, the actin filaments in the simulations were assumed to be about an order of magnitude shorter than in experiments. This effectively reduced the separation between the actin and PEG length-scales in the simulations compared to the experiments, potentially making the morphological changes in simulations less sharp, as compared to the experiments. The model configuration space is built from a choice of model specific interaction parameters (see Methods). These parameters were tuned to recover the same qualitative morphological behavior as observed in experiment. In future work, these parameters could be systematically tuned to investigate the effects of changing model interaction strengths which would likely recover the more abrupt morphological difference observed experimentally.

An important consideration is that PEG is not a protein. As shown before, for PEG $R_g \sim MW^\nu$ with $\nu = \frac{3}{5}$. For proteins, the radius of gyration has been modeled as being proportional to $MW^{1/3}$ [51] which, for the analytical case of two spheres, gives a scaling equivalent to that of a spherical colloid, $\zeta = -\frac{1}{2}$ as before. Fitting experimental data for a combination of different proteins [53], as well as other experimental work [54] gives a scaling of $R_g \sim MW^{2/5}$. This exponent $\nu = \frac{2}{5}$ give a $U^*$ that scales with $q$ as $q^{-1/2}$ just as with spherical colloids. We note that the value of $\nu$ for which $\zeta$ becomes positive is $\nu = \frac{1}{2}$. Proteins, with their scaling exponent at $\nu = \frac{2}{5}$ are just below this threshold. Despite not recovering the sign change we saw analytically for the case of proteins; we believe that further exploration of fixed-mass depletion is of interest in a cellular context [3,10]. It is clear that the

Asakura-Oosawa model doesn't provide a complete description. For instance, ATP-dependent activity in the cell can enhance diffusion [55]. Further, one could include electrostatic, and van der Waals interactions. Finally, the Asakura-Oosawa model treats depletants as an ideal gas, an assumption which is only valid in the limit of low depletant concentrations. We note that the PEG concentrations in our study are more dilute than in cellular contexts. As such, we believe that the effect could be much stronger in cells due to macromolecular crowding [50,51]. We further note that the diameter of a typical crowding protein is ~ 5 nm, but the ranges of sizes varies within the cytosolic context [3,52]. The effect of depletant size polydispersity could additionally be taken into consideration. Future studies can focus on incorporating these effects to refine the form of $U^*$, and determine if positive scaling is present within the cellular context when considering fixed-mass depletion. Furthermore, it would be interesting to determine if the magnitude and sign of $\zeta$ could be actively controlled to assist in dynamic intracellular organization.

**Fixed-mass depletion in relation to material properties.** Confocal microscopy indicated that to the left of the blue line (cf. Fig. 2a), actin filaments coexist with bundles. To the right of this line, the entire network appears to be incorporated into bundles. The difference in our test configurations, $\Gamma_{6k}$ and $\Gamma_{20k}$ states, is supported by DLS analysis (cf. Fig. 3). The fitted relaxation times for each condition are significantly different and match our observations in confocal microscopy. Namely, as the molecular weight of the PEG polymers increases, the relaxation time increases because a greater fraction of the network is bundled. The stretched exponent also decreases, indicating a greater heterogeneity in the distribution of bundle sizes. The $\Gamma_{20k}$ condition, by nature of being forced into the bundled regime is most likely to be kinetically arrested [56]. What would have been free diffusion is limited by inter-bundle steric interactions. In contrast, the control condition (no PEG) is most freely diffusing; $\tau_f$ is small and the intensity pattern decorrelates quickly. Since the $\Gamma_{6k}$ morphology exists as a mixture of the control condition and the $\Gamma_{20k}$ conditions it combines features of both the kinetic arrest and the free diffusion modes. Given this, one would expect that the stretched exponent would be smaller for the $\Gamma_{6k}$ condition. However, the opposite is observed – where the stretched exponent is smallest for the $\Gamma_{20k}$ condition. The reasoning for this is not clear within the context of this analysis. Further quantitative work with our scattering data is difficult when using Mie scattering assumptions, such as spherical particle shapes, which cannot be applied to actin filaments. Studies have shown that dynamic light scattering data can also be analyzed by changing the dynamic structure factor of the fit to a model suited for semiflexible polymers [57–59]. The decay rate associated with this semiflexible polymer dynamic structure factor depends on the persistence length of the polymer, which can be used as a fit parameter. With this type of analysis, we report values for the persistence lengths of the $\Gamma_{6k}$ condition to the $\Gamma_{20k}$ condition, which are statistically indistinguishable from each other yet show increases in respective mean values (see Figure S8).

Previous work has characterized bundle size properties, such as the diameter [26]. The FRET analysis we presented here provides an additional measure on bundle geometry, namely the intra-bundle spacing between actin filaments. With PEG as a bundling agent, we expect the filament bundles to be in the fully coupled regime [60]. Our analysis gives us a measure on the distance associated with fully coupled actin filaments bundled by depletion forces. This could be extended to measure intra-bundle distance for different actin bundling proteins. Scaling out, the physics that limits the extent of bundle diameter is still very much an open question. When well-mixed, we observe that actin filaments bundle in a polydisperse way (see Figure S9), with bundles forming simultaneously across the full spatial extent of the system. Given this, there are two different mechanisms to consider. The first is the mechanism by which a free actin filament gets incorporated into a neighboring bundle. The other is the interaction between two mature bundles. For the formation of a single bundle theoretical arguments have been proposed in terms of chirality [61], packing defects [62], and counterion repulsion [63]. Per the bundle-bundle interaction, arguments have been presented pointing to the interplay between surface-surface interactions and macroscopic hydrodynamic forces in the system [24]. The gels in our study exhibit the polydisperse incorporation of filaments into bundles across the full spatial extent of the system, both in the model and experimentally. The polydisperse morphology of bundles seen in the model only includes surface-surface interactions, suggesting that the hydrodynamic forces aren't as dominant in determining subsequent network morphology. Future work could investigate the limitations on bundle diameter more systematically via similar measurement techniques to those presented in the main text, and supplemental information. On the scale of an individual bundle, electron microscopy and FRET measurements give bounds on the number of actin filaments incorporated into a bundle for states and further, these modalities inform that bundle diameter and intra-bundle spacing are equivalent across these morphological regimes (see supplemental information). FRET also gives insight into the notion of effective crosslinking by parameterizing how close filaments within a bundle are to one another. One might expect to uncover physical signatures between different mechanisms that drive bundling based on extending the FRET methods outlined herein. We note that further characterization of similar gels has been performed, studying the water dynamics using 2DIR spectroscopy [64].

When looking at the bulk properties of the gel, the storage modulus for $\Gamma_{20k}$ is larger than for $\Gamma_{6k}$ across all replicates, indicating a difference in gel stiffness. The $\Gamma_{20k}$ condition has a longer diffusion lifetime as shown by dynamic light scattering measurements, perhaps due to the increased stiffness of the bulk material. We also note that we observe stronger gels than previously reported [26]. We suspect this is due in part to the omission of gelsolin in the current work which is known to truncate the length of actin filaments and could impact bulk network properties as a result. Further, the current study uses 50mM KCl vs 1M KCl. Using higher concentrations could impact filament-filament interactions and reduce gel stiffness.

Previous work from Hosek and Tang [24] predicted that the critical concentration of bundling depends on PEG molecular weight and presented experimental results to validate their claim. In addition, our current study expands on this work by exploring the system below the critical concentration, namely the weakly-bundled regimes to give a more complete characterization of the phase space for these networks. Depletion of rods has been studied in other systems, such as with viruses [65–68]. The phase spaces resulting in these systems are quite different from the observed behavior in our study [67]. We attribute this to the fact that we are working in more dilute samples which may give simpler phase behavior. The scaling laws presented here focus on the pair potential minimum. Still, the integrated effect of the attractions is typically what determines the phase behavior. If the minimum of the fixed-mass depletion pair potential gets deeper with depletant molecular weight, then its second virial coefficient will also become more negative (i.e. stronger integrated attractions) since its range grows. The fixed number density pair-potential minimum gets shallower with increasing molecular weight, but its range gets larger. It's not clear whether the second virial will become positive or negative. However, what is clear is that the integrated attractions of the fixed-mass case will become increasingly stronger than the fixed-number-density case as the molecular weight grows. This is an important distinction since both the minimum and the range matter for thermodynamics. In this work, we haven't sought to extend our analysis of fixed-mass depletion to the expected phase behavior. We've only started to get at this experimentally (cf. Figure 2a). This would be a potential avenue for future work.

## Conclusion

In this study we explored how the strength of depletion interaction potential $U^*$ varies when the total depletant mass is held fixed, a.k.a. fixed-mass depletion. We answered this question by studying networks of reconstituted semiflexible actin *in silico*, and *in vitro*. Our work gives insight into the biologically relevant phenomena of fixed-mass depletion and provides a more complete characterization of the depletion interaction, which is necessary to better understand the intracellular organization of the cytoskeleton [3]. We've demonstrated actin morphology can be changed in a fixed-mass context, by changing the molecular weight of the PEG molecules that act as depletants in the system. The ability to dynamically change actin morphology via steric interactions in a crowded but dynamic environment is reminiscent of known mechanisms in living cells, such as within condensates [69] or macromolecular crowding [50]. Future work could explore other areas of our configuration space using similar characterization techniques as done in this work.

Our work also provides a foundation for deriving semiflexible-polymer-based biomimetic materials that might embody some of life's dynamic ability for accomplishing mechanical tasks. Our identification of a morphological change from a weakly-bundled state, to a strongly bundled state offers insight to explain the dynamic properties of cell-based actin manipulation, and perhaps a pathway towards harnessing this protein in artificial contexts. We have demonstrated control over actin network morphology and mechanics by steric interactions alone under the constraint of fixed mass of the depletants. A better mechanistic understanding of fixed-mass depletion could allow for *in situ* control over material properties by varying the size distribution of a closed system with fixed depletant mass (e.g. aggregation or lysis of depletants). Future work could leverage a similar numerical modeling framework to test for reversibility [24] in network morphology to help motivate studies into reversible soft matter materials [70]. Further developing this type of control could lead to novel strategies in developing bio-inspired materials that mimic cells' ability to dynamically mechanically adapt and respond to stimuli [24,70].

## Methods

**Actin Preparation.** Actin was purified from rabbit psoas skeletal muscle from Pel-Freeze using a GE Superdex 200 Increase HiScale 16/40 column and stored at −80 °C in G-Buffer (2 mM tris-hydrochloride pH 8.0, 0.2 mM disodium adenosine triphosphate (ATP), 0.2 mM calcium chloride, 0.2 mM dithiothreitol). All protein stocks were clarified of aggregated proteins at 100 000g for five minutes upon thawing and used within seven days. The G-actin concentration in the supernatant was determined by measuring the solution absorbance at 290 nm with a Nanodrop 2000 (ThermoScientific, Wilmington, DE, USA) and using an extinction coefficient of 26 600 M$^{-1}$ cm$^{-1}$.

**Confocal Microscopy.** Samples were prepared to yield a final buffer concentration of 20 mM imidazole pH 7.4, 50 mM potassium chloride (KCl), 2 mM magnesium chloride ($MgCl_2$), 1 mM dithiothreitol, 0.1 mM ATP, 1 mM trolox, 2 mM protocatechuic acid (PCA), and 0.1 mM protocatechuic 3,4-dioxygenase (PCD). The polyethylene glycol, KCl, imidazole, dithiothreitol, and $MgCl_2$ were purchased from Sigma-Aldrich. The adenosine triphosphate and trolox were purchased from Fisher Scientific. PCA was purchased from the HWI Group. PCD was purchased from Sigma Aldrich.

Glass flow cells were prepared by sonication of individual slides in water for 5 minutes, followed by blow-drying with nitrogen. These slides were then placed in a base piranha solution of five parts DI Water, one part 30% hydrogen peroxide, one part 30% ammonium hydroxide for thirty minutes at 80 °C. These slides were then sonicated again for 5 minutes, blow-dried with nitrogen, and stored in isopropanol until use. Thick coverslips and thin slides were attached by means of melting Parafilm with pre-cut chambers, which are then treated with potassium hydroxide for ten minutes to activate hydroxyl groups on the glass surface and then passivated with 0.2 mg/ml poly-l-lysine–g—polyethylene-glycol from Nanosoft Polymers in a humid environment [71].

We used an Olympus FV1000 motorized inverted IX81 microscope suite, with instrument computer running FV10-ASW software version 4.2b software, to image actin networks using laser-scanning confocal microscopy. Actin filaments were labelled with rhodamine–phalloidin on a one-to-one molar ratio and excited with 543 nm wavelength laser light. Each sample of actin was prepared once and imaged in three random, well-separated locations.

Each *z*-stack taken was processed using ImageJ. The image was opened and a maximum *z*-projection across 21 μm through the bulk of the image was produced. For each confocal *z*-stack both the degree of bundling and the mesh size were algorithmically determined (See SI).

**Rheology.** Experiments were performed on a TA Discovery HR 20 rheometer fitted with either a 20 mm, 2° stainless steel cone geometry or an 8 mm parallel plate. Master buffer was prepared according to the steps described in the Confocal Microscopy section, else the PCA, PCD, trolox. Master buffer was mixed by pipette with deionized water and PEG molecules of appropriate molecular weight and in the desired concentration. Actin was then added and mixed gently by pipette before adding to a 1:1 molar ratio, actin:phalloidin. After combining and mixing with dried phalloidin, the sample was pipetted onto the rheometer with a total volume of 75 μL. Throughout all experiments, a temperature-controlled Peltier plate maintained the temperature at 25°C, and a solvent trap was utilized to prevent evaporation during data acquisition. Immediately after loading the sample, a time sweep was started to monitor the evolution of the shear moduli over time. The time sweeps were performed at 2% strain and a frequency of 1 rad/s. These parameters were selected based on strain sweeps performed at 1 rad/s (Figure S7) and frequency sweeps performed at 2% strain (Figure S8).

**Dynamic Light Scattering.** Actin and master buffer were prepared according to the steps described in the Confocal Microscopy section, else the PCA, PCD, Trolox. Phalloidin in methanol was dried using compressed nitrogen gas and added to the sample solution in a 1:1 molar ratio with the actin to stabilize the filaments and prevent depolymerization. 45 μL were imaged in a Malvern Zetasizer Nano ZS instrument for five runs for each sample. The refractive index of the master buffer was found to be 1.334, and the refractive index of actin was found to be 1.3343.

Particles in a fluid are known to scatter incident light. The diffusion of the suspended particles changes the reported intensity at each angle measured; therefore, using a directed laser, it is possible to use the intensity measurement across multiple temporal decades to understand the Brownian dynamics of the particles in solution. This is done by measuring the diffraction field at a given time $\tau$, and then re-measuring the sample at some later time [72].

$$G_2(\tau) - 1 = G_1(\tau) = \sigma^2 \left( e^{-\left(\frac{\tau}{\tau_f}\right)^\beta} \right)^2 \tag{8}$$

For spherical particles, this is usually accomplished using the Stokes-Einstein equation, which directly gives the hydrodynamics radius. However, for non-spherical particles this calculation is no longer valid, and a quantitative comparison requires a calculation of the stretched exponent $\beta$.

**FRET.** G-actin, Atto 488, and AlexaFluor 555 maleimide dye (Thermo Fisher) were thawed and 1 mL G-buffer 1x was prepared and protein stock was clarified of aggregates as described above. The volume of dye needed to achieve an excess of 10x dye to protein was calculated. The dye was added to G-actin solution and mixed thoroughly by pipette prior to 2 hours of incubation at room temperature, allowing the dye to bind actin cystines. During incubation, Princeton 20 centrispin columns are hydrated with 650 µL of G-buffer, allowing the resin to swell for 30 minutes. After incubation, the hydrated columns were placed in a centrifuge and spun at 700 g for 2 minutes to remove excess G-buffer. The dyed protein solution was then added to the hydrated column by pipette, being careful to avoid the edges of the column while pipetting. The columns were then placed in a centrifuge and spun at 700 g for 2 minutes, where the gel filtration column effectively separates the actin from the free dye. Concentrations were then measured via Nanodrop. Labelled G-actin was aliquoted, flash-frozen with liquid nitrogen, and stored at $-80\,°C$ until experimentation.

Sample chambers are constructed with piranha etched #1.5 glass microscope coverslips and 2 mm thick silicon gaskets with a 5 mm diameter hole punched out. Prior to assembly the slides and gaskets are treated in Hellmanex solution at 80 °C for 20 minutes to ensure adhesion between the two components. 30 minutes prior to addition of protein samples, the chambers are passivated with 1 mg/ml bovine serum albumate (BSA) – purchased via Sigma Aldrich - to prevent interactions between the actin and the glass slide. During slide passivation the sample is mixed via a multi-step procedure. The first step is preparing donor seed filaments where half of the total sample, deionized water, and master buffer are mixed with the volume of donor-labelled (Atto 488) actin monomers by pipette. This polymerizes the donor filaments, which are then stabilized by adding to 1:1 phalloidin. The donor seed filaments are then added to the other half of the total sample, deionized water, and master buffer along with PEG molecules, acceptor-labelled (AlexaFluor 555) actin, and unlabelled actin monomers and mixed gently by pipette. This is then added to 1:1 phalloidin to stabilize the filaments. The BSA in the sample well is removed by washing 5 times with master buffer. After the final wash, the master buffer is removed and 20 µL of combined protein sample is added to the sample well. The final concentration for all samples is as follows. [Atto 488 actin] = 0.1 µM, [AlexaFluor 555 actin] = 1.0 µM, [Total Actin] = 12 µM.

The sample was loaded onto a home-built time-correlated single photon counting (TCSPC) confocal fluorescence microscope. The microscope utilizes a 486 nm picosecond laser with a 50 MHz repetition rate and laser power set to 50 µW via neutral density attenuation. For all samples, the laser focus was place at a depth of 5 µm from the base of the microscope slide using a Mad City Laboratories piezo stage and translated via micrometer adjustment in *x* and *y* dimensions to navigate to different spatial locations. Emitted photons were collected using a 1.45 NA, 100x magnification microscope objective and routed through a pinhole and 511/10 bandpass emission filter toward a Hamamatsu GaAsP photomultiplier tube. Photomultiplier output pulses were then amplified and counted with a Becker and Hickl (BH) TCSPC computer card. For each replicate, 3 random, well-separated regions were imaged to incorporate intra-sample variation in our results.

Lifetime data was analysed using BH SPCImage software where decay-matrix calculations were performed to generate a distribution of fluorescence lifetimes within an image. The quantity $\tau_{DA}$ is determined from this distribution in fluorescence lifetimes. The quantity $\tau_D$ was determined with a control study where the distribution of fluorescence lifetimes was measured for actin filaments that were only labelled with the donor. The average donor-to-acceptor distance was then calculated using equations (1) and (2).

**Simulation Parameters.** The actin bead-chains and have a uniform length of 160 nm and bead diameter ($\sigma$) of 8 nm, and the diameter of the PEG spheres is varied from 0.25 $\sigma$ to 0.75 $\sigma$. All distances in the simulations are scaled by $\sigma$, and all times by the time ($\tau$) it takes for a bead in an actin bead-chain to diffuse across a distance $\sigma$ The diffusion coefficient for a freely diffusing actin monomer is estimated to be $10^1 \,\mu m^2 s^{-1}$, which gives $\tau = 0.1s$ [73]. All energies in the model system are scaled by $K_b T$. The simulation box size is set to 75 $\sigma$ by 75 $\sigma$ by 10 $\sigma$ across all simulations, and periodic boundary conditions are enforced.

To understand how the (1) concentration and (2) molecular weight of the PEG particles influenced the actin bundling, we vary (1) the number of PEG particles from 4000 to 10000 while keeping the simulation box volume unchanged and (2) PEG particle radius from 2 nm to 6 nm, respectively. To convert from PEG radius of gyration ($R_g$) to molecular weight ($M$), we assumed the classic result from Flory theory, $R_g \propto M^{\frac{3}{5}}$. This gives molecular weights of 6k, 20k, and 37.4 k for radii of 0.25 $\sigma$, 0.5 $\sigma$, and 0.75 $\sigma$ respectively. The PEG concentration is varied from 4000 $\frac{V_\rho}{V_{box}}$, 5000 $\frac{V_\rho}{V_{box}}$, 6000 $\frac{V_\rho}{V_{box}}$, and 10000 $\frac{V_\rho}{V_{box}}$, where $V_\rho = \frac{4}{3}\pi r^3$, $r$ representing the effective radius of and $V_{box}$ is the volume of the simulation box which is 56250 $\sigma^3$. We explore 12 different systems to determine the effects of PEG concentration and effective molecular weight. The number of actin filaments is held constant at 200 for all 12 systems.

For the Lennard Jones pairwise interactions, $\epsilon_{PEG-PEG}$ is set to 1.25 for attractive interactions and 1.0 for repulsive interactions; $\epsilon_{actin-actin}$ is set to 0.3 for attractive interactions and 1.0 for repulsive interactions. We encode only repulsive interactions between the PEG beads and the actin strands, and $\epsilon_{PEG-actin}$ is equal to 2.0.

## Author Contributions

James Clarke is the main contributor in this work. Lauren Melcher, Anne Crowell, and Francis Cavanna are recognized as having contributed equally to this work. Lauren Melcher led the development and analysis of the simulation and its results with Moumita Das. Anne Crowell helped perform all rheological experiments and wrote much of the analysis and methods related to rheology. Francis Cavanna performed the preparation and confocal imaging of actin networks, as well as the preparation for DLS imaging. Justin Houser helped to prototype FRET experiments and provided direct assistance in all FRET data acquisition. Allison Green performed the DLS measurements, fitted the stretched exponential and relaxation time parameters, and provided insight and analysis with Delia Milliron. Kristin Graham helped with acceptor-donor labeling for actin-FRET measurements, along with experimental planning for all FRET data acquisition. Tom Truskett provided insights into the nature of morphological differences, as well as limitations on bundle diameter. Adrianne Rosales was instrumental in conceptually designing rheological experiments and provided consistent feedback and insight as the experiments progressed. Jeanne Stachowiak was instrumental in planning FRET experiments. José Alvarado planned most of the inter-lab experimentation, assisted in the interpretation of all the results, and provided direct feedback on the structure and material of this paper.

## Conflicts of interest

There are no conflicts to declare.

## Acknowledgements

This research was primarily supported by the National Science Foundation through the Center for Dynamics and Control of Materials: an NSF MRSEC under Cooperative Agreement No. DMR-1720595 and DMR-2308817 with additional support from the Welch Foundation (F-1848 and F-1696) and NSF DMR-2144380. The authors acknowledge the use of facilities and instrumentation supported by the National Science Foundation through the Center for Dynamics and Control of Materials: an NSF MRSEC under Cooperative Agreement No. DMR-1720595. We also acknowledge the Texas Materials Institute for the use of facilities and instrumentation. We recognize Carlos Baiz and Xiaobing Chen for insightful discussions. We also acknowledge Vernita Gordon for use of her confocal microscope, as well as Raluca Gearba-Dolocan for assistance with electron microscopy. José Alvarado and Moumita Das would like to thank the Isaac Newton Institute for Mathematical Sciences, Cambridge, for support and hospitality during the programme - New statistical physics in living matter: non equilibrium states under adaptive control – where work on this paper was undertaken. This work was supported by EPSRC grant No. EP/K032208/1.

## Notes and references

# Supplemental Material
**Mesh Size Determining Algorithm**

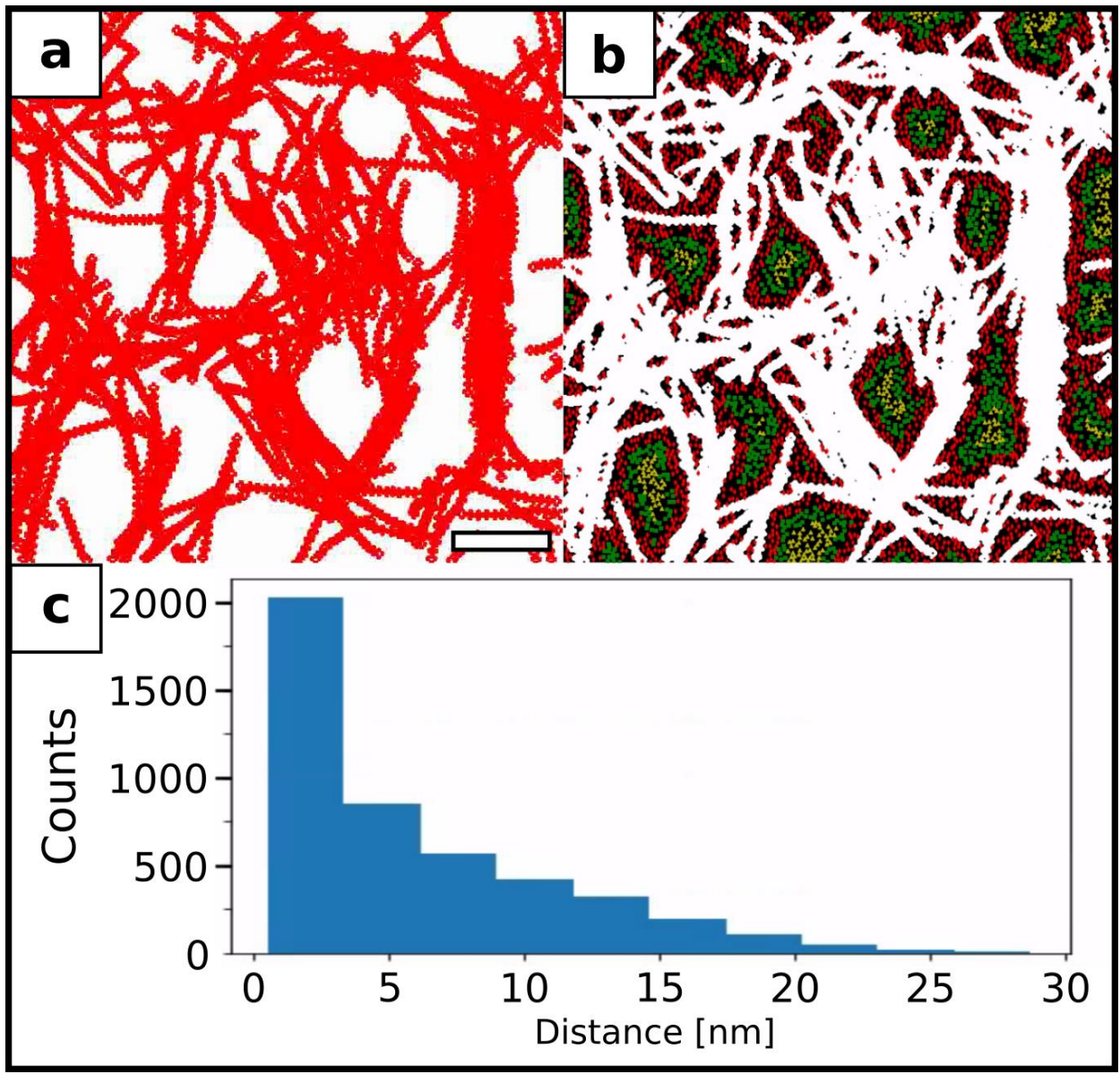

Figure S1: (a) Real-space representation of numerical simulations taken from $r_1$ replicate 1. Actin (shown in red) is being depleted by PEG (not shown). (b) Overlay image detailing the meshsize algorithm behavior. White pixels - actin filaments, just as spatially represented in (a). Red diamonds - points with a calculated distance < 8 nm to nearest filament. Green squares - points with a calculated distance between 8 nm and 15 nm to nearest filament. Yellow triangles - points with a calculated distance > 15 nm to nearest filament. Overlay Scalebar = 100 nm. (c) Distribution across all points measured in (b).

The algorithm drops 10000 random points onto the image into the gaps (pixel value 0) and calculates the shortest distance to the nearest actin filament (pixel value 1). This result is as expected, we look at the distribution of these points and see that there are more points in the red group with fewer in green, and even fewer in yellow. The mean of this distribution gives the mesh size. As the mesh size increases, we would see that the number of green and yellow points would increase relative to the red points, shifting the mean towards higher values, vis versa we'd see that the yellow points, and even green points would start to disappear as the mesh size decreases, resulting in a smaller value for the mean in that case. For these mesh size calculations, we follow previously developed methods [10,11].

**PEG Clustering**

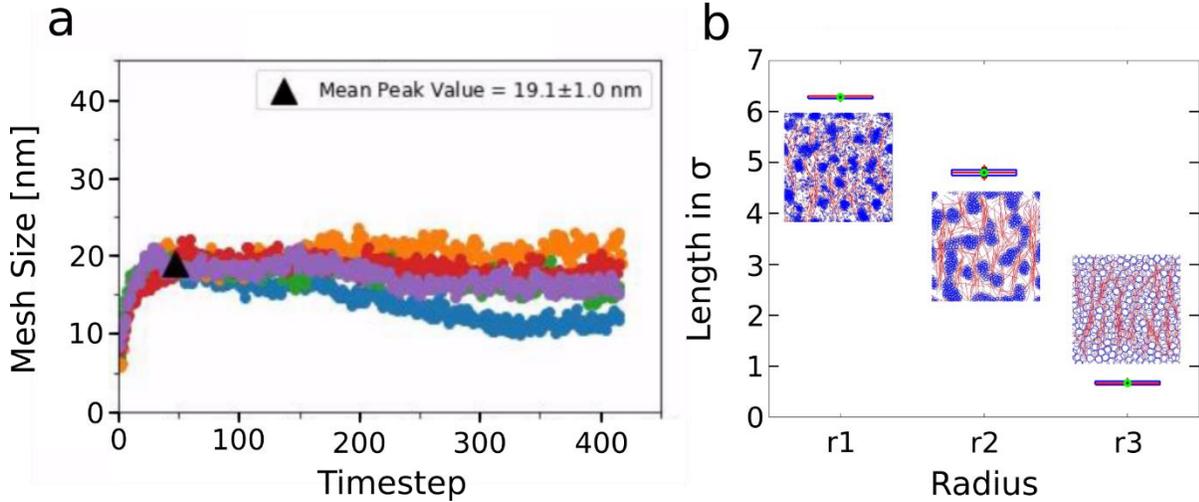

Figure S2: (a) Representative simulation trace of mesh size. Black triangle indicates where the mesh size was determined (as reported in main text Figure 2b). (b) All pairwise comparisons of distributions for $r1$, $r2$, $r3$, return statistically significant differences (****) when compared via t-test.

**Number Density view of confocal data**

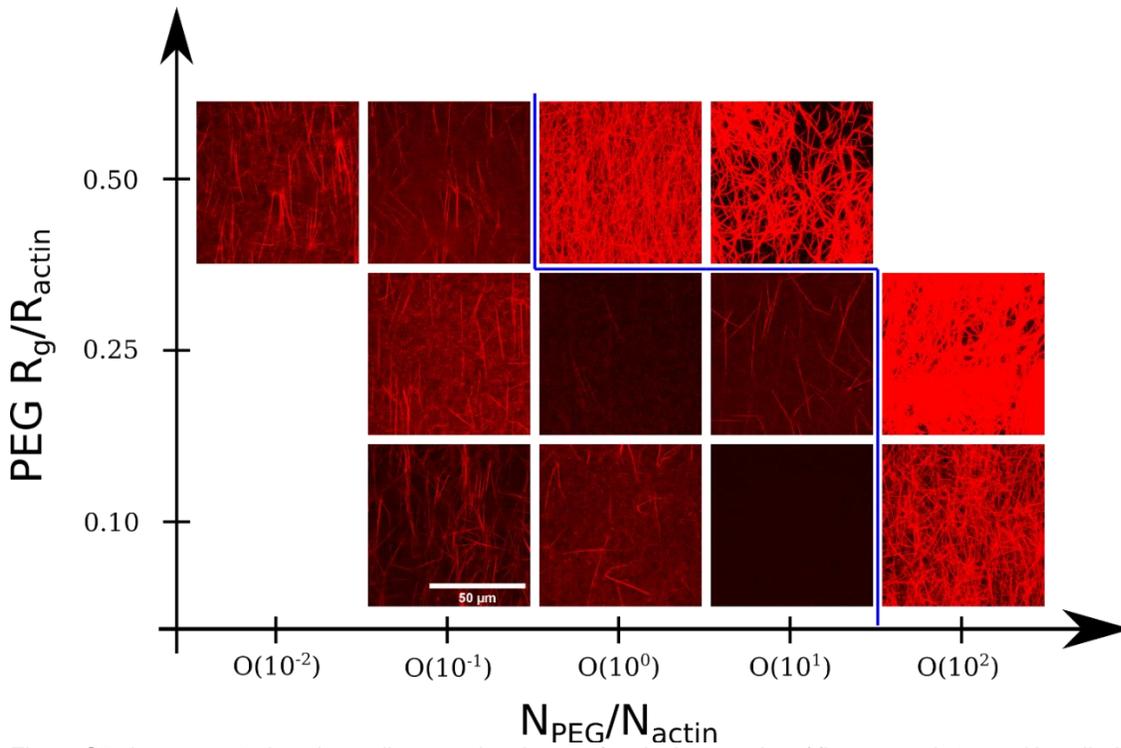

Figure S3: A representative phase diagram showing confocal micrographs of fluorescently tagged bundled actin. [actin] = 12 µM with various concentrations (shown as a ratio of number density) of PEG on the x-axis and the size ratio of

PEG to actin on the y-axis corresponding to a total of 36 distinct samples, giving 3 replicates for each condition. Scale bar = 40 $\mu m$.

## Degree of Bundling Algorithm

The procedures to determine each are as follows. Z-stacks are normalized and subsequently binarized with an Otsu filter. Binarized volumes are then skeletonized using Scikit-Image's skeletonize routine. The number of nodes is then extracted from the skeleton. The number of nodes in the skeleton are taken as proxy for how effective the PEG molecules are in bundling the actin network. For a filamentous network, one would expect the number of skeleton nodes to be large – as one would need more skeletons to describe all unique contours in the actin network. Conversely, for a strongly bundled network, one would expect the number of skeleton nodes to be small. We then take the inverse of the number of nodes to give us our metric for the degree of bundling.

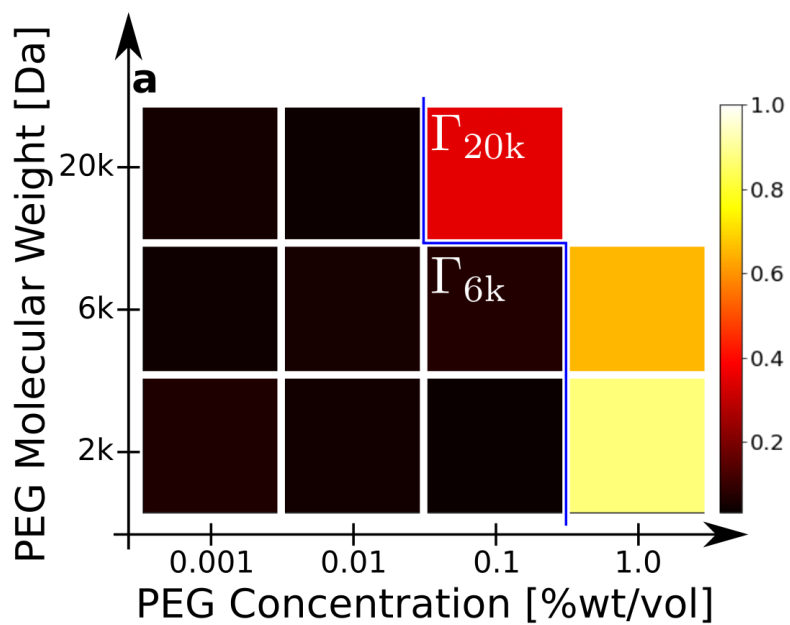

Figure S4: Heat maps representing the degree of bundling of the phase space depicted in main text Figure 2a, as found by our Degree of Bundling skeletonizing algorithm. Each value corresponds to N=3 measurements per sample condition.

## Dynamic Light Scattering (DLS)

**Raw DLS Curves.**

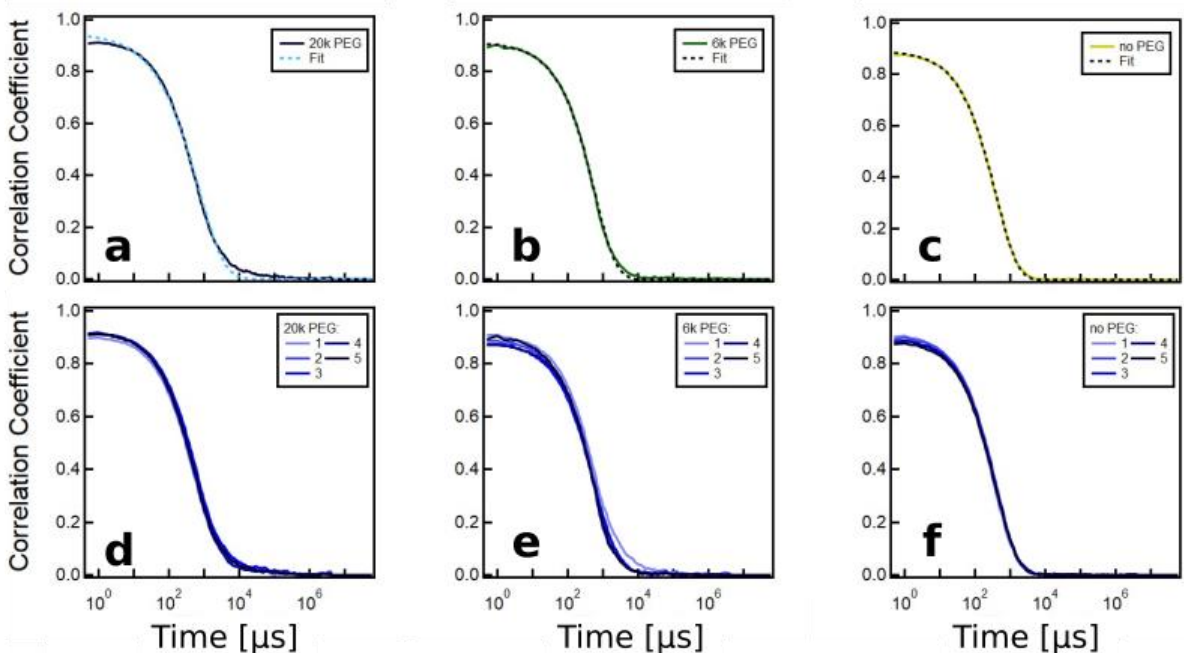

Figure S5: Fitting of correlation function data from dynamic light scattering (DLS) measurements, against $\Gamma_{20k}$ and $\Gamma_{6k}$ and a control of [actin] = 12 µM with no PEG. Panels a-c) contains a sample fitting of one of the five replicates for each condition. Panels d-f) represent the total intensity correlation functions measured for each experimental condition.

## Rheology

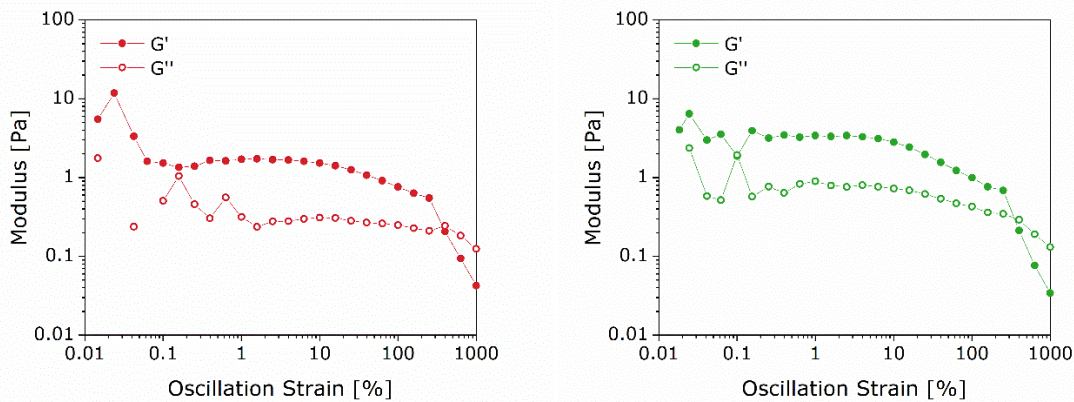

Figure S6: Strain sweeps, performed at an angular frequency of 1 rad/s, were used to identify appropriate strain settings for time sweeps and frequency sweeps. The storage moduli ($G'$) and loss moduli ($G''$) are shown for $\Gamma_{6k}$ (red) and $\Gamma_{20k}$ (green). In both cases, a strain of 2% was in the linear region and therefore was applied for the other measurements.

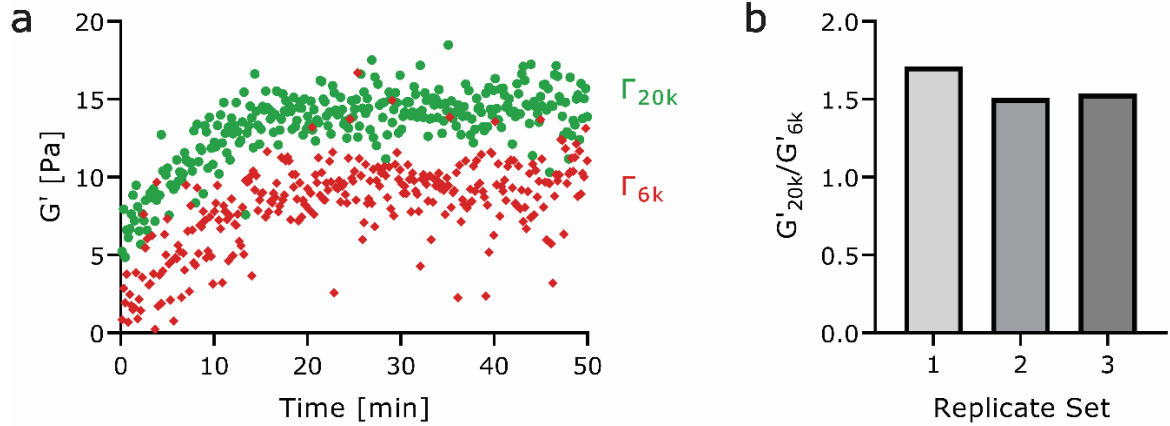

Figure S7: Rheometry data was also collected using an 8 mm parallel plate geometry. Representative time sweeps are shown in (a), and comparisons of the storage moduli for the two formulations across three replicates are shown in (b). The data for $\Gamma_{6k}$ (red) and $\Gamma_{20k}$ (green) revealed similar trends as those seen with the 20 mm cone geometry: higher storage moduli were reached for $\Gamma_{20k}$. However, the values of the measured moduli were higher than those measured with the cone, likely due to more significant effects of instrument inertia and evaporation when using the 8 mm parallel plate. The effect of instrument inertia on the data collected with the 8 mm parallel plate geometry was evidenced by phase angles greater than 170 degrees. As a result, the measured moduli are higher than the actual values associated with the materials. In addition, significant evaporation was noted for measurements with the 8 mm parallel plate geometry. Since solvent traps are not commercially available for this geometry, a damp wipe was placed near the sample and the sample area was covered, though significant evaporation still occurred. By comparison, very little (if any) evaporation occurred using the 20 mm cone with a solvent trap. For these reasons, the measurements with the cone geometry are considered more reflective of the material properties and are shown in the main text.

**Estimation of Persistence Length.** In the regime of semiflexible polymers, the dynamic structure factor is given by,

$$G_2(\tau) - 1 = G_1(\tau) = G_1(0) \left[ -\frac{\Gamma\left(\frac{1}{4}\right)}{3\pi} \left[ \frac{k_B T}{4\pi\eta} \left( \frac{5}{6} - \ln(q2a) \right) \right]^{3/4} \frac{q^2 t^{3/4}}{L_p^{1/4}} \right] \tag{1}$$

,

where $\Gamma$ is the gamma function, $k_B$ is the Boltzmann constant, $T$ is temperature, $\eta$ is the kinematic viscosity, $q$ is the scattering vector, $a$ is the mean hydrodynamic radius of the meshwork and $L_p$ is the persistence length. To perform the fits, the hydrodynamic radius of the system was first set by tuning the persistence length of the control to match the expected value for actin meshwork [1]. Subsequently, DLS curves were fit using Eq. 1 directly to extract the persistence lengths.

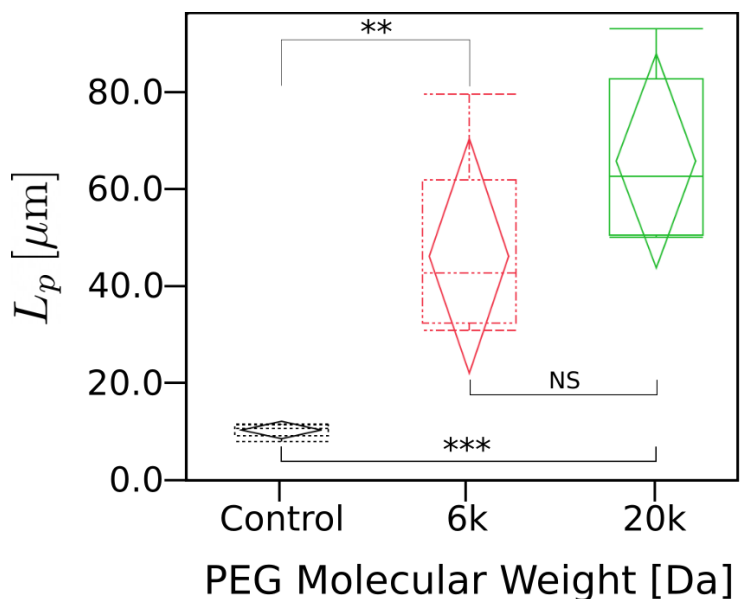

Figure S8: Distributions of measured persistence lengths, $L_p$ for dynamic light scattering experiments. Control corresponds to a filamentous actin meshwork in the absence of PEG molecules. The $\Gamma_{6k}$ and $\Gamma_{20k}$ samples are statistically indistinguishable, per this analysis. This is denoted by "NS". Both $\Gamma_{6k}$ and $\Gamma_{20k}$ are statistically differing from the filamentous actin meshwork.

## Electron Microscopy (EM)

**Electron Microscopy Methods.** Experiments were performed on a Thermo Scientific Scios 2 DualBeam through the Texas Materials Institute (TMI). Sample holders were constructed using silicon wafers purchased from Montco Silicon Technologies Inc. The wafers were cut with a diamond scribe to match the geometry of the EM sample stubs. The silicon wafers were cleaned using a rinse of deionized water, followed by blow drying with a clean nitrogen line. The scribed silicon substrates were adhered to the sample stubs using double-sided carbon dots adhesive stickers. The silicon wafers were connected to the sample stub via aluminum tape to ensure proper electrical contact through the sample.

Master buffer was prepared according to the steps described in the Confocal Microscopy section, else the PCA, PCD, Trolox. Master buffer was mixed by pipette with deionized water and PEG molecules of appropriate molecular weight and in the desired concentration. Actin was then added and mixed gently by pipette. The sample was then added to a 1:1 molar ratio, actin:phalloidin and subsequently mixed. The sample was then pipetted onto the silicon wafer in a manner such as to ensure a flat film of solution across the surface as much as possible.

Samples were transported to TMI and loaded into an EMS sputter coater. The sputter coater was run with the following settings: Current = 40mA, Deposition time: 45 seconds, Species = Au/Pt: 60/40. The sputter coated sample was then transferred directly to the Scios 2 Dual Beam for imaging. For Scanning Electron Microscope (SEM) depositions the following settings were used: Carbon deposition – 15x10x0.05 µm at 5kV, 3.2nA. For SEM imaging the following settings were used: OptiPlan Mode 2kV, 1.2nA, Working Distance 7mm. FIB processing was carried out as follows: Step 1: Pt deposition 15x10x1µm at 30kV, 1nA; Step 2: Regular Cleaning Cross-section 17x10x10µm at 30kV, 15nA; Step 3: Cleaning Cross-section 18x5x12µm at 30kV, 1nA. Stage tilt was at 52 degrees for all cross-sectional cuts.

Image analysis was performed using Python where an active contours method was implemented to measure critical dimensions of actin bundles as visualized by SEM on fresh cross-sectional FIB cut surfaces.

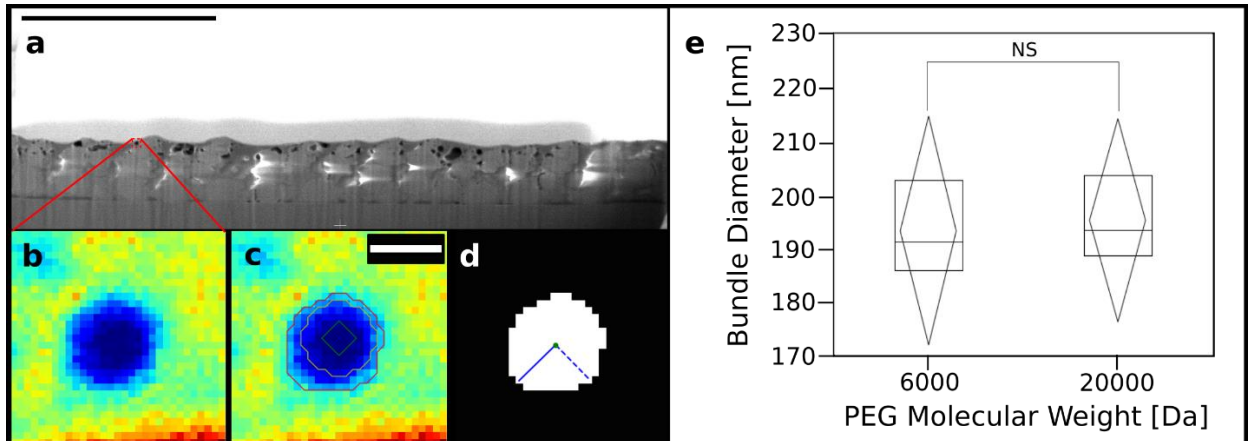

Figure S9: (a) Cross-sectional SEM micrograph of PEG bundled actin network for Γ6k. Red dashed box indicates the subset of the image corresponding to panels b-d. Scale bar = 5 μm (b) Pre-processed subset of SEM micrograph with one actin bundle. (c) Active contour overlay of SEM micrograph subset seen in panel b. Green, yellow, red represent the first, 5th, and final contours, respectively. Top right white colored scale bar = 100 nm. (d) Resultant area considered for critical dimension measurement of bundle diameter with the centroid (green dot), major axis (solid blue line) & minor axis (dashed blue line) displayed. (e) Quartile box plots with confidence diamonds of mean bundle diameter size in nanometers of network extracted from cross-sectional electron micrographs for the $\Gamma_{6k}$ and $\Gamma_{20k}$ conditions. The two populations are statistically indistinguishable, denoted by "NS" in the panel.

## Bounds for number of filaments in a bundle

The results of the FRET analysis also enable us to estimate the number of filaments incorporated into the bundles given the measured cross-sectional size of bundles from the electron microscopy experiments, the measured intra-bundle distance, and the diameter of an actin filament [2]. If we assume a hexagonal packing geometry, we get a measured number of filaments of $N_{FRET} = 217$. It is worth mentioning that our protocol for SEM imaging requires desiccation of the actin network, which may lead to a coalescence of bundles and therefore an overestimation of the bundle thickness. Bundles incorporating 100s of filaments have been observed in drosophila bristles in the presence of different crosslinking proteins, forked proteins and fascin [3–5]. In comparison, while our reported values of persistence length via DLS are similar to earlier work with fascin crosslinks [6], results from the semiflexible DLS analysis indicate only less than ten filaments incorporated into bundles for $\Gamma_{6k}$ and $\Gamma_{20k}$ sample conditions. This is determined by calculating the bending modulus from the persistence length (cf. Fig. 4) and applying the known scaling for fully coupled bundles [7]. In some part, this can be related to the coexistence of filaments and bundles in the $\Gamma_{6k}$ case. The underestimate is less clear in the $\Gamma_{20k}$ case. In the semiflexible polymer DLS analysis we calibrate one value of the hydrodynamic radius $a$ as a tuning parameter on the control (no PEG) system. In doing this, there isn't any change in $a$ for the $\Gamma_{6k}$ and $\Gamma_{20k}$ fits. Further, with increasing the molecular weight of PEG solutions in water, the viscosity is shown to increase slightly [8]. We use one value of kinematic viscosity for all fits which leads to an additional source of error [9]. Together, the results from the semiflexible polymer DLS and electron microscopy give bounds on the number of filaments that are incorporated into bundles for the $\Gamma_{6k}$ and $\Gamma_{20k}$ sample conditions.